\documentclass[graybox]{svmult}

\usepackage{mathptmx}       
\usepackage{helvet}         
\usepackage{courier}        
\usepackage{type1cm}        
\usepackage{makeidx}         
\usepackage{graphicx}        
\usepackage{multicol}        
\usepackage[bottom]{footmisc}
\usepackage{floatrow}
\usepackage{subfigure}

\usepackage{gensymb}

\newcommand{\be}{\begin{equation}}
\newcommand{\ee}{\end{equation}}
\newcommand{\bea}{\begin{eqnarray}} 
\newcommand{\eea}{\end{eqnarray}}
\newcommand{\ft}[2]{{\textstyle\frac{#1}{#2}}}

\def\pd{\ensuremath{\partial}}
 \def\cL{{\cal L}}

\makeindex            

\begin{document}

\title*{CMB Anisotropies by Collapsing Textures}

\author{Kepa Sousa and Jon Urrestilla}

\institute{Kepa Sousa \at School of Engineering and Science, Jacobs University Bremen, \email{ksousa@jacobs-university.de}
\and Jon Urrestilla \at Department of Theoretical Physics, University of the Basque Country UPV/EHU, \email{jon.urrestilla@ehu.es}}

\maketitle

\vspace{-0.3cm}
\abstract{CMB photons passing through a collapsing texture knot receive an energy shift, creating characteristic cold and hot spots on the sky.  We calculate the anisotropy pattern produced by collapsing texture knots of arbitrary shape.  The texture dynamics are solved numerically on a Minkowski background. }
\vspace{-0.3cm}
\section{Introduction}
\label{sec:1}
\vspace{-0.3cm}

Textures are an unstable type of topological defect which  are generically formed whenever  there is a complete spontaneous breaking of a non-abelian global symmetry, what can be easily implemented in the context  of Grand Unified Theories   \cite{Turok:1990gw}.
 Unlike other topological defects textures are unstable, and collapse roughly at the speed of light as soon as they enter the horizon. When the defect size falls below the symmetry breaking scale the topological charge is no longer conserved and the texture decays into the vacuum (\emph{unwinding}). The integrated Sachs-Wolfe effect causes the CMB photons passing near the texture to be typically red-shifted or blue-shifted, leaving characteristic hot and cold spots in the cosmic background. The interest on textures increased after Cruz \emph{et al.} and Feeney \emph{et al.} \cite{Cruz:2007pe,Feeney:2012jf}  considered the texture model as one of the most plausible hypothesis to explain the CMB anomaly known as the \emph{Cold Spot}.

\begin{figure}[t!]
\floatbox[{\capbeside\thisfloatsetup{capbesideposition={left,top},capbesidewidth=0.4\textwidth}}]{figure}[\FBwidth]
{\caption{Prediction of the time evolution of the temperature fluctuation at the center of the anisotropy by different methods: field theory simulations (solid line), self-similar collapse (dashed line)(\ref{TurokAnisotropy})\label{fig:timeProfile}}}
{\includegraphics[width=0.5\textwidth]{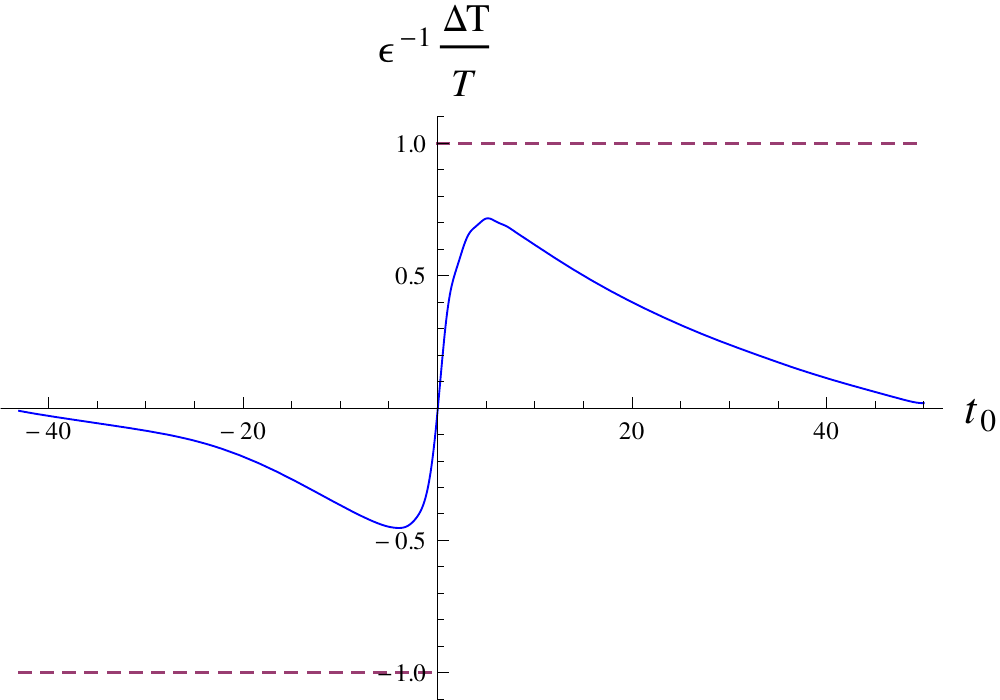}}
\vspace{-0.6cm}
\end{figure}

These  analyses rely on the existing predictions for the anisotropy pattern produced by global textures. In particular they use a very idealized analytical result obtained by Turok \emph{et al.} in \cite{Turok:1990gw}, who studied the simplest model admitting  texture solutions, the $\mathrm{O}(4)-$model, which is characterized by the Lagrangian
\be
\cL = \ft12 \pd_\mu  \phi_a \, \pd^\mu \phi_a - \lambda (\phi_a \phi_a -\eta^2 )^2, \qquad  a =1,2,3, 4.
\label{lagrangian}
\ee
This model describes the dynamics of four real scalar fields $\phi_a$, with their interactions given by a mexican-hat type  potential,  where $\lambda$ is the self coupling  and  $\eta$ the expectation value, which determines the symmetry breaking scale. 
 The analytical solution found by Turok  \emph{et al.} relies on the  non-$\sigma$ model approximation to solve the dynamics (i.e. $\phi_a \phi_a=\eta^2$ is assumed at all times), which breaks down at the unwinding event, and describes  spherically symmetric texture collapsing in a self-similar way. The anisotropy pattern produced by such a texture is given by
\be
\frac{\Delta T}{T}(r,t_0) = \frac{t_0}{(2 r^2  + t_0^2)^{1/2}} \; \varepsilon, \qquad \varepsilon \equiv  8 \pi^2 G \eta^2  , 
\label{TurokAnisotropy}
\ee
where $r$ is the impact parameter of the photon respect to the center of the texture, and $t_0$ the time at which it is closest to the texture, with  $t_0=0$ being the time at unwinding.  
 Such a solution has been known for a long time to be unlikely to occur in a cosmological context \cite{Borrill:1992pm}. Actually, the authors of   \cite{Cruz:2007pe,Feeney:2012jf} truncated the radial profile (\ref{TurokAnisotropy}) and matched it with a Gaussian at its half-maximum because it is known not to be valid for large values of the impact parameter where it has a very slow decay as $r^{-1}$,  leading to an unrealistically large spot.  Moreover, the amount of red-shift or blue-shift received by the photons crossing the center of the texture is  the same independently of the time of crossing $t_0$, (dashed line in Fig. \ref{fig:timeProfile}), even at very early times when the texture size is large, and thus the energy density is very diluted. 

The purpose of the present work is to provide a more realistic prediction of the anisotropy pattern left by a random texture, without making any assumption about the initial configuration of the texture, through numerical simulations of the full dynamics of a texture configuration in the $\mathrm{O}(4)-$model.

\vspace{-0.3cm}
\section{Numerical methods}
\vspace{-0.3cm}
In order to characterize the anisotropy produced by textures with arbitrary shape we proceed as in  \cite{Borrill:1994uh}. To simplify the dynamics it is  assumed that the gravitational field produced by the texture is small, which implies that the texture can be evolved on the unperturbed background and the CMB photons travel along the unperturbed geodesics. The dynamics of the $O(4)-$model are solved evolving numerically a discretized version of the equations of motion for Minkowski background on a lattice of $96^3$ grid points. The lattice has periodic boundary conditions, and the grid spacing is $\Delta x =  \sqrt{2/\lambda \eta^2}$. 
These results can be extended to the cosmological case as long as all the length-scales and time-scales involved in the simulation are small compared with the size of the horizon $H^{-1}$. We have evolved 1300  random initial configurations with a correlation length of $36$ lattice spacings, and during a time interval of $96 \Delta x \,\cdot c$, so that the boundary effects can  be ignored. The corresponding anisotropy pattern is calculated only for those initial configurations leading to isolated unwinding events which happen away from the start and the end of the evolution (33 in total), so that photons have time enough to cross the complete texture. In a Minkowski background the Sachs-Wolfe formula can be solved explicitly in terms of the energy momentum tensor of the texture configuration. The anisotropy is calculated using an approximation of this solution valid  for anisotropies covering small angular scales (see \cite{Borrill:1994uh}). 

\vspace{-0.3cm}
\section{Summary of the results and conclusions}
\label{Conclusions}
\vspace{-0.3cm}

In order to compare our results with the self-similar solution (\ref{TurokAnisotropy}), we have measured the fractional temperature change for photons crossing the center of the texture for each of the 33 initial configurations, and then we have averaged over the whole ensemble. The result is represented by the solid line in Fig. \ref{fig:timeProfile}. The plot shows how our numerical simulation resolves the unwinding event, which lasts about  $\delta t \sim 10 \  \lambda^{-1/2} \eta^{-1}$, in contrast with the analytic solution in \cite{Turok:1990gw} which has a step-like behavior (dashed line). Moreover, we can also see how the brightness of the spot decays at early and late times, implying that textures are only observable during a finite interval around the unwinding event. 
This might have important consequences for the Bayesian analyses in \cite{Cruz:2007pe,Feeney:2012jf}, which requires an estimate of the number of cold and hot spots  due to textures which can be observed in the sky at a given time, and of a given angular size. In particular, as the size of the texture configuration grows with time as we go far from the unwinding event \cite{Turok:1990gw,Borrill:1994uh}, we expect the angular scale distribution of spots to decay faster for large spots than the estimate made in \cite{Turok:1990gw}. We have also recovered  results in \cite{Borrill:1994uh} which show that the average cold and hot spots produced by random textures are significantly less pronounced than the Turok solution ($20-50 \%$): 
\be
\frac{\Delta T}{T}|_{max} = (+0.77  \pm 0.21) \varepsilon, \qquad \frac{\Delta T}{T}|_{min} = (-0.49  \pm 0.13) \varepsilon,
\ee
which  is specially relevant in order to estimate the symmetry breaking scale, as it can be extracted from the spot brightness \cite{Cruz:2007pe,Feeney:2012jf}. Moreover, the reduction on both the spot brightness and the number of spots might hinder distinguishing the signature of an unwinding event on the CMB temperature from a large gaussian fluctuation. The analysis of the CMB polarization would provide a useful method to discriminate between those two possible scenarios \cite{Vielva:2010vn}.

\begin{figure}[t!]
\includegraphics[width=0.44\textwidth]{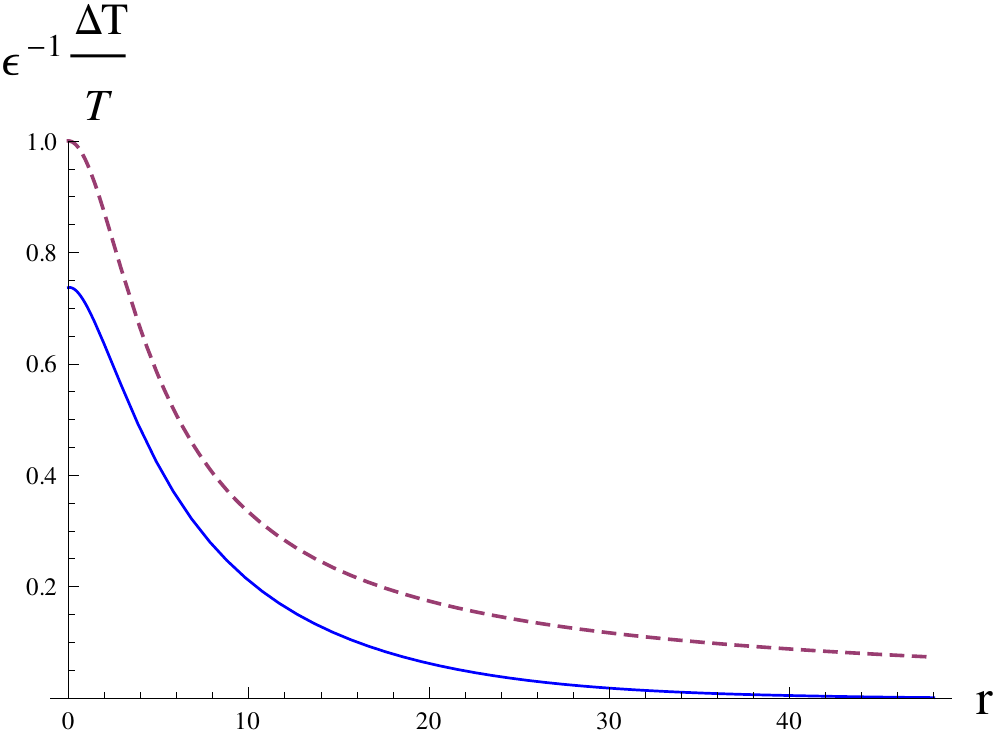}
\hspace{0.3cm}
\includegraphics[width=0.44\textwidth]{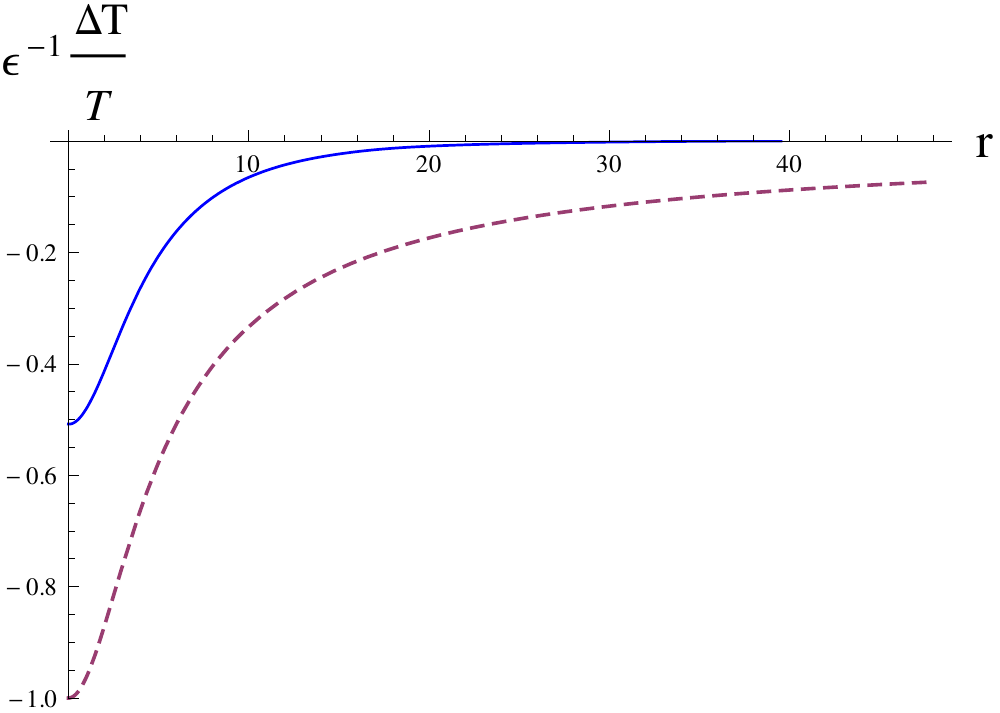}
\caption{ Averaged radial profiles of a \emph{hot spot} (left)   and a \emph{cold spot} (right) at  maximum brightness. The continuous lines represent our results and the dashed line corresponds to eq.  (\ref{TurokAnisotropy}). 
} \label{fig:averageProfiles}
\vspace{-0.4cm}
\end{figure}

In order to characterize the anisotropy pattern we have obtained  the radial profiles of a cold and a hot spot  at  maximum brightness. Since  textures become spherical close to the time of the unwinding \cite{Press:1989yh},  after averaging over the whole ensemble of initial conditions, we have also averaged the profile over the azimuth angle. 
The result is represented   in Fig. \ref{fig:averageProfiles} (continuous lines). In these plots we can see again the differences between the maxima and minima of the calculated profiles and the Turok solution  (dashed line). In addition, it is also evident that the profiles we found are significantly more localized than the analytic solution. These results, together with the expected corrections to the angular scale distribution of observed spots suggest  a revision of the analyses done by Cruz \emph{et al.} and Feeney \emph{et al.}   \cite{Cruz:2007pe,Feeney:2012jf}.

The present study is a first approach to improve the existing predictions for the anisotropy pattern produced by a collapsing texture. Future work involves repeating these simulations in a larger lattice in order to reduce the boundary effects, and to study the dependence of the profiles on the photon emission and reception times. In addition we intend to evolve the texture configurations in a Friedman-Robertson-Walker metric in order to characterize the effect of the expansion of the universe on the anisotropy pattern produced.

\vspace{-0.4cm}

\begin{acknowledgement}
We are grateful to M. Cruz, and E. Martinez for very useful discussions. KS  acknowledges support within
the framework of the Deutsche Forschungsgemeinschaft (DFG) Research Training Group 1620 Models of gravity, and thanks the Department of Theoretical Physics at the University of the Basque Country for its hospitality. JU acknowledges financial support from the Basque Government (IT-559-10), the Spanish Ministry (FPA2009-10612), and
the Spanish Consolider-Ingenio 2010 Programme CPAN (CSD2007-00042). 
\end{acknowledgement}
\vspace{-0.9cm}


\begin{thebibliography}{99.}%
%
%


\bibitem{Borrill:1992pm}
  J.~Borrill, E.~J.~Copeland and A.~R.~Liddle,
  Phys.\ Rev.\ D {\bf 46} , p. 524.


\bibitem{Borrill:1994uh}
  J.~Borrill, E.~J.~Copeland, A.~R.~Liddle, A.~Stebbins and S.~Veeraraghavan,
  Phys.\ Rev.\ D {\bf 50}, p. 2469

\bibitem{Cruz:2007pe}
  M.~Cruz, N.~Turok, P.~Vielva, E.~Martinez-Gonzalez and M.~Hobson,
  Science {\bf 318}, p. 1612


\bibitem{Feeney:2012jf}
  S.~M.~Feeney, M.~C.~Johnson, D.~J.~Mortlock and H.~V.~Peiris,
  Phys.\ Rev.\ Lett.\  {\bf 108}, p. 241301
%

\bibitem{Press:1989yh}
  W.~H.~Press, B.~S.~Ryden and D.~N.~Spergel,
  Astrophys.\ J.\  {\bf 347}, p. 590.


%


\bibitem{Vielva:2010vn}
  P.~Vielva, E.~Martinez-Gonzalez, M.~Cruz, R.~Barreiro and M.~Tucci,
MNRAS {\bf 410 (1)}, p. 33. 


\bibitem{Turok:1990gw}
  N.~Turok and D.~Spergel,
  Phys.\ Rev.\ Lett.\  {\bf 64}, p. 2736.


%

\end{thebibliography}
\end{document}